\documentclass[aps,prb,reprint,superscriptaddress,floatfix]{revtex4-2}

\usepackage{amsmath,amssymb}
\usepackage{graphicx}
\usepackage{xcolor}
\usepackage{bm}          
\usepackage{siunitx}     
\usepackage{booktabs}    
\usepackage{hyperref}
\hypersetup{colorlinks=true, linkcolor=blue, citecolor=blue, urlcolor=blue}
\usepackage{placeins}

\begin{document}

\title{Large spin splitting metallic altermagnets from machine-learned design rules}

\author{Ali Sufyan}
\affiliation{NanoLund and Division of Mathematical Physics, Department of Physics, Lund University, SE-221 00 Lund, Sweden}
\affiliation{Wallenberg Initiative Materials Science for Sustainability, Department of Physics, Lund University, SE-221 00 Lund, Sweden}

\author{Brahim Marfoua}
\affiliation{Department of Physics, Chemistry and Biology, Link\"oping University, SE-581 83 Link\"oping, Sweden}
\author{J. Andreas Larsson}
\affiliation{Applied Physics, Division of Materials Science, Department of Engineering Sciences and Mathematics, Lule\aa\ University of Technology, SE-971 87 Lule\aa\, Sweden}
\affiliation{Wallenberg Initiative Materials Science for Sustainability, Lule\aa\ University of Technology, SE-971 87 Lule\aa\, Sweden}
\author{Rickard Armiento}
\affiliation{Department of Physics, Chemistry and Biology, Link\"oping University, SE-581 83 Link\"oping, Sweden}
\author{Erik van Loon}
\affiliation{NanoLund and Division of Mathematical Physics, Department of Physics, Lund University, SE-221 00 Lund, Sweden}
\affiliation{Wallenberg Initiative Materials Science for Sustainability, Department of Physics, Lund University, SE-221 00 Lund, Sweden}
\email{erik.van_loon@fysik.lu.se}

\date{\today}

\begin{abstract}
Altermagnets combine compensated magnetic order with momentum-dependent spin splitting, providing spin-polarized electronic states without a net magnetization. Metallic $d$-wave altermagnets are particularly promising because they can support time-reversal-odd spin currents in linear response, yet experimentally validated bulk realizations remain scarce. Here, we use high-throughput density-functional-theory data to relate the magnitude of altermagnetic band splitting to compositional, structural, and DFT-derived magnetic descriptors. An interpretable gradient-boosted model identifies two candidate-prioritization criteria: compact unit cells and magnetic sublattices that sustain sizable local moments. Guided by these trends, we screen tetragonal $A_2XY$ Heusler compounds in space group $P4/mmm$. Among 307 structures, symmetry identifies 169 altermagnetic arrangements, of which 154 remain metallic altermagnets in DFT. Sixteen realize an altermagnetic collinear ground state, of which 15 are dynamically and mechanically stable and 10 also lie on or below the calculated thermodynamic hull. Six candidates exceed the CrSb splitting obtained under the same computational protocol, led by Co$_2$AlSc ($\Delta_\mathrm{max}=2.24$~eV) and Fe$_2$AlGe ($2.07$~eV). A Julliere-model estimate gives a tunneling magnetoresistance of up to $356\%$ at the Fermi level for Co$_2$TiZn. These results identify a chemically tunable family of metallic $d$-wave altermagnets and demonstrate how interpretable machine learning can guide targeted first-principles searches.
\end{abstract}

\maketitle

\section{Introduction}

Altermagnets constitute a third class of collinear magnetism, distinct from both ferromagnets and conventional antiferromagnets: in the nonrelativistic limit, their opposite-spin sublattices are related by a rotational crystal symmetry rather than by a pure translation or inversion, so the net magnetization vanishes while the electronic bands acquire a momentum-dependent spin splitting of nonrelativistic origin~\cite{Smejkal2022a,Smejkal2022b,Jungwirth2026, devaraj2026altermagnetism}. This combination makes altermagnets attractive for spintronics. Their compensated magnetic order strongly suppresses macroscopic stray fields and may facilitate dense device integration, while their spin-split bands enable ferromagnet-like transport phenomena, including giant magnetoresistance, tunneling magnetoresistance~\cite{SmejkalTMR2022}, spin-splitter currents~\cite{GonzalezHernandez2021}, and anomalous transport responses~\cite{Smejkal2020}. The availability and form of these responses are constrained by symmetry. Spin-group symmetry classifies the momentum-space spin splitting according to the minimum number of symmetry-protected spin-degenerate nodal surfaces, two, four, or six, corresponding to $d$-, $g$-, and $i$-wave order~\cite{Smejkal2022a,Jungwirth2026}. All three classes can support spin Hall and tunneling magnetoresistive responses when permitted by the magnetic symmetry and transport geometry. By contrast, the nonrelativistic spin-splitter response occurs at linear order only for $d$-wave order; for $g$- and $i$-wave order, the corresponding leading responses occur at third and fifth order in the electric field, respectively

Because the splitting is set by exchange rather than spin-orbit coupling, it can reach the scale of electron volts without relying on heavy elements. Metallic $d$-wave altermagnets with large exchange-driven spin splittings therefore represent attractive candidates for spin-transport studies. Several altermagnets have been established in experiments, but there is no general consensus on a bulk material that combines metallicity, unambiguous $d$-wave order, and a large directly resolved spin splitting. Spin-split bands have been resolved by photoemission in semiconducting
MnTe~\cite{Krempasky2024,lee2024broken,Osumi2024}, in metallic CrSb thin films and crystals~\cite{Reimers2024,Ding2024,zeng2024observation,Yang2025crsb}, and in the intercalated dichalcogenide CoNb$_4$Se$_8$~\cite{Regmi2025,dale2024relativistic,Jungwirth2026}, all three of which are $g$-wave. In the $d$-wave class, inelastic neutron scattering has established two-dimensional altermagnetism in La$_2$O$_3$Mn$_2$Se$_2$, but this material is a correlated
insulator~\cite{wei20252,asai2026realization}.

The layered oxychalcogenides KV$_2$Se$_2$O and Rb$_{1-\delta}$V$_2$Te$_2$O were reported as metallic
room-temperature $d$-wave altermagnets on the basis of
photoemission~\cite{Jiang2025,Zhang2025svl}. Subsequent neutron
diffraction, however, identified bulk antiferromagnetic order
in both KV$_2$Se$_2$O, with $T_\mathrm{N}\approx400$~K, and
Rb$_{1-\delta}$V$_2$Te$_2$O, with
$T_\mathrm{N}\approx337$~K~\cite{Sun2025,xie2026g}, which challenges the interpretation of these materials as bulk $d$-wave altermagnets.
However, surface- or layer-resolved
altermagnetism may still account for the spin splitting observed by
surface-sensitive probes~\cite{Chakraborty2026}.

For rutile RuO$_2$, transport and photoemission studies have reported altermagnetic signatures~\cite{Feng2022,Bai2023,Bose2022,Fedchenko2024}, whereas muon spin rotation finds a nonmagnetic ground state~\cite{hiraishi2024nonmagnetic}; its bulk magnetic state therefore remains disputed. This discrepancy may reflect sensitivity to disorder and strain~\cite{Smolyanyuk2024}: dynamical mean-field theory places RuO$_2$ near both the paramagnetic-to-altermagnetic phase boundary and the itinerant-to-localized crossover, and predicts that $\sim0.5\%$ compressive strain can stabilize altermagnetism~\cite{Park2026}. Mn$_5$Si$_3$ is another metallic $d$-wave candidate, supported by anomalous Hall and Nernst measurements but without direct observation of a large band splitting~\cite{Reichlova2024,Badura2025}. Thus, consensus on the identification of a bulk metallic $d$-wave altermagnet with a large, directly resolved spin splitting remains to be established.

Beyond these experimental examples, computational surveys based on spin-group symmetry~\cite{Smejkal2022a}, high-throughput spin-polarized DFT~\cite{Guo2023,Sufyan2026}, and DFT combined with embedded dynamical mean-field theory~\cite{Wan2025} consistently find that symmetry-allowed altermagnetism is considerably more common than its metallic realization, with most candidates being insulating or semiconducting. For Fermi-surface spin transport, the principal bottlenecks are therefore metallicity and the magnitude of the splitting near the Fermi level. Screening over known magnets yields splittings predominantly below a few hundred meV~\cite{Sufyan2026,Guo2023}; for example, a library of magnetically intercalated dichalcogenides identifies seventeen $g$-wave candidates with Fermi-level splittings of order $100$~meV~\cite{DayRoberts2026}. Data-driven and symmetry-aided approaches are expanding the search beyond databases of known magnetic structures~\cite{cui2026data,zhou2026magnetic}. Machine-learning models have likewise been used for symmetry classification, candidate ranking, and splitting prediction~\cite{Bhattarai2025,Gao2025,Jiang2026}, with recent searches reporting calculated splittings above $1.5$~eV, including $1.88$~eV in RbMn$_2$Te$_2$O~\cite{Jiang2026}, although values obtained using different computational protocols should not be compared as absolute records. Complementary chemical principles relate the splitting to the anisotropy and relative orientation of local coordination motifs~\cite{YuanZunger2023,Wei2024}, while the absence of high-order rotations within a same-spin sublattice favors $d$-wave symmetry~\cite{Wei2024,Fender2025}. These rules explain the symmetry and anisotropy of the splitting but do not quantitatively rank its magnitude across broad chemical spaces.

Here we use machine learning as a tool for knowledge extraction to identify physically interpretable trends and test them through targeted first-principles screening. A gradient-boosted model trained on 180 DFT-labeled structures from our MAGNDATA survey~\cite{Sufyan2026} identifies compact unit cells and large local moments per magnetic atom as the descriptors most strongly associated with large spin splitting. These amplitude-related trends complement motif geometry, which determines the symmetry and momentum dependence of the splitting, and motivate a survey of tetragonal $A_2XY$ Heusler compounds ($P4/mmm$). Starting with 307 candidates from the Alexandria database~\cite{Cavignac2026Alexandria}, symmetry analysis of the Wyckoff positions of the magnetic atoms gives 169 altermagnetic candidates, of which 154 are confirmed as metallic $d$-wave altermagnets by DFT; three further compounds relax to an uncompensated ferrimagnetic state and are excluded. In terms of stability, 16 materials satisfy $\Delta E_{\mathrm{AM-GS}}\leq0.1$~meV/f.u, 15 are dynamically and mechanically stable and 10 lie on or below the GGA+$U$ convex hull. We find six altermagnets with a spin splitting larger than CrSb, led by Co$_2$AlSc ($2.24$~eV) and Fe$_2$AlGe ($2.07$~eV). We further evaluate their directional Fermi-surface polarization and tunneling response, and calculate the intrinsic spin Hall conductivity for four representatives. These results identify $A_2XY$ Heuslers as a chemically tunable platform for metallic $d$-wave altermagnetism and demonstrate that interpretable trends from a small data set can guide searches for exceptionally large spin splitting.

\section{Computational Details}
\label{sec:comp}

All density-functional-theory (DFT) calculations were performed using the Vienna \textit{Ab initio} Simulation Package (VASP)~\cite{Kresse1996,Kresse1999} and the Perdew--Burke--Ernzerhof (PBE) exchange--correlation functional~\cite{Perdew1996}. The electron--ion interactions were described using projector-augmented-wave (PAW) datasets~\cite{Blochl1994,Kresse1999} from the VASP \texttt{potpaw\_PBE.54} library (version 54, released in September 2015). The element-specific POTCAR choices followed the Materials Project \texttt{PBE\_54} mapping documented by Kingsbury \textit{et al.}~\cite{kingsbury2022performance}. Hubbard corrections were applied within the rotationally invariant Dudarev scheme~\cite{Dudarev1998} (\texttt{LDAUTYPE}\,=\,2), using element-specific effective interaction parameters $U_\mathrm{eff}=U-J$. The $U_\mathrm{eff}$ values were adopted from our previous high-throughput survey~\cite{Sufyan2026}. Plane-wave cutoffs of 400~eV and 520~eV were used for structural relaxations and subsequent static calculations, respectively, with \texttt{PREC}=\texttt{Accurate}. Reciprocal space was sampled using $\Gamma$-centered meshes generated with target spacings of $0.40$~\AA$^{-1}$ for relaxations and $0.25$~\AA$^{-1}$ for static calculations. Gaussian smearing (\texttt{ISMEAR}=0) with $\sigma=0.05$~eV was applied throughout. Electronic self-consistency was converged to $10^{-5}$~eV during relaxation and $10^{-6}$~eV for the static and band-structure calculations. Ionic relaxation was terminated when all residual forces were below $0.05$~eV,\AA$^{-1}$ (\texttt{EDIFFG} =$-0.05$), with the cell shape, volume, and internal coordinates relaxed simultaneously (\texttt{ISIF}=3). Structural relaxations, magnetic-energy comparisons, and calculations of the nonrelativistic altermagnetic splitting were performed without spin--orbit coupling. Phonon stability was assessed using the finite-displacement method implemented in \textsc{phonopy}~\cite{Togo2023}, with $2\times2\times2$ supercells. Mechanical stability was evaluated from the calculated elastic tensors using the Born criteria for tetragonal crystals. Thermodynamic stability was assessed within GGA+$U$ using separate sets of competing phases obtained from the Materials Project and Alexandria databases \cite{horton2025accelerated, jain2013commentary, Cavignac2026Alexandria}, including relevant elemental, binary, ternary, and same-composition cubic ($Fm\bar{3}m$) Heusler competitors. Altermagnetic classifications were obtained with \textsc{amcheck}~\cite{amcheck} and independently verified using the symmetry operations returned by \textsc{spglib}~\cite{spglib}. Relativistic magnetic space groups were determined with \textsc{spglib} using vector magnetic moments. Maximally localized Wannier functions were constructed with \textsc{Wannier90}~\cite{Pizzi2020} from VASP wavefunctions, including the spin matrix elements. Spin--orbit coupling was included in these relativistic transport calculations. The resulting Wannier Hamiltonians were symmetrized onto the corresponding magnetic space groups, and the intrinsic spin Hall and anomalous Hall conductivities were evaluated with WannierBerri~\cite{Tsirkin2021} on adaptively refined $k$-point meshes.

\begin{figure}[t]
 \centering
\includegraphics[width=\columnwidth, trim=0pt 10pt 0pt 0pt]{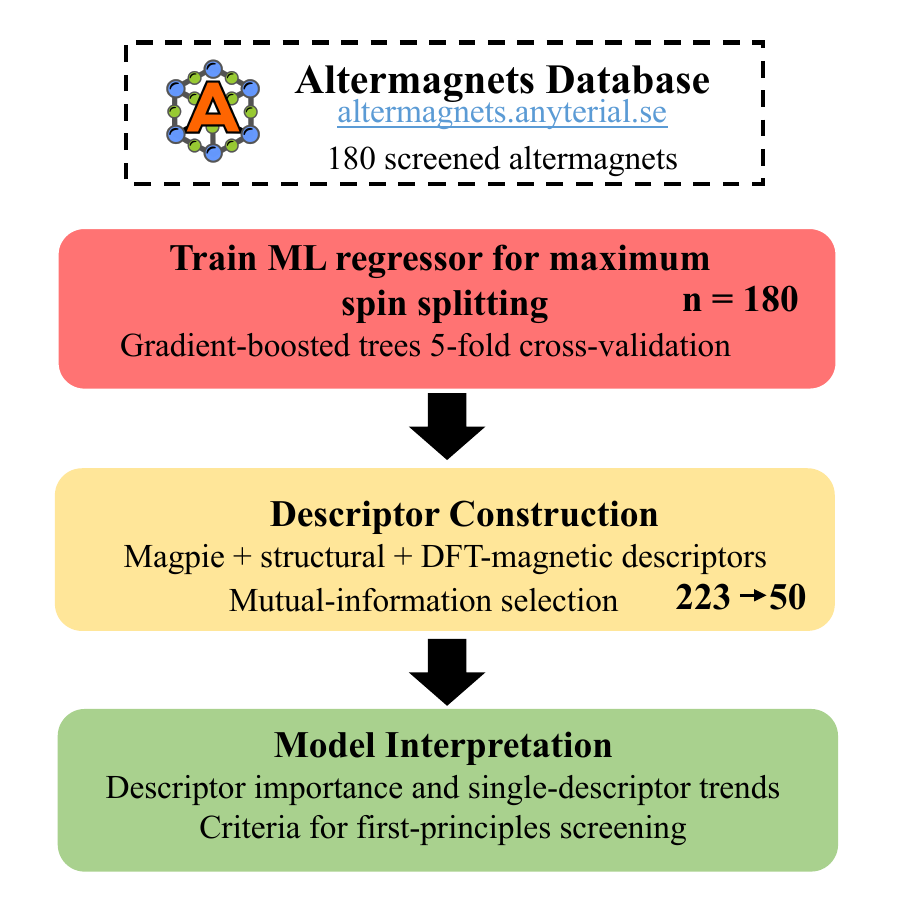}
\caption{Machine-learning workflow used in this work. DFT-labelled
  materials from the previous MAGNDATA screening are represented by
  compositional, structural, and DFT-derived magnetic descriptors.
  Mutual-information selection reduces the initial set of 223
  descriptors to 50, which are used to train a gradient-boosted
  regressor for the maximum altermagnetic band splitting
  $\Delta_\mathrm{max}$. Model interpretation is then used to formulate
  criteria for first-principles candidate selection.
}
\label{fig:ml_workflow}
\end{figure}

The regression data set comprised the 180 DFT-labelled materials of our previous high-throughput study of the MAGNDATA database~\cite{Sufyan2026,gallego2016magndata1,gallego2016magndata2}. For each material, 223 descriptors were constructed with \textsc{pymatgen}~\cite{Ong2013}: compositional statistics following the Magpie framework~\cite{Ward2016} (means, standard deviations, and ranges of electronegativity, atomic radius, Mendeleev number, ionization energy, covalent radius, and $s/p/d$ valence-electron counts), structural descriptors (lattice parameters $a$, $b$, $c$, the $c/a$ ratio, $N_\mathrm{atoms}$, and space-group number), and DFT-derived magnetic quantities read from the relaxed altermagnetic calculation (per-atom magnetic moments and their statistics, number of magnetic atoms, Hubbard $U$ values, collinear spin angle, and band gap). Because the splitting distribution spans more than two orders of magnitude with a strongly right-skewed profile (mean 0.15~eV, median 0.13~eV), all models were trained on the transformed target $y=\ln\left(\Delta_\mathrm{max}/\mathrm{eV}+0.01\right)$, and predictions were transformed back to eV before evaluation. Fifty descriptors were retained by mutual-information scoring against $y$; this selection was performed once on the full data set, whereas descriptor standardization was fitted within each cross-validation fold. Four model architectures were compared by fivefold cross-validation: a baseline gradient-boosted regressor (GBR) with fixed hyperparameters, a random forest, a tuned GBR
whose hyperparameters were optimized by \texttt{RandomizedSearchCV} (80 iterations, fivefold inner loop, maximizing $R^2$ on the transformed target), and a two-step model that first classifies materials as large or small splitters and then applies separate regressors. Among models whose overall cross-validated $R^2$ lay within 0.05 of the best, the tuned GBR was retained because it gave the highest cross-validated $R^2$ on the large-splitting subset ($\Delta_\mathrm{max}>0.3$~eV), the scientifically most relevant regime. Feature attributions for the final model were
computed with SHapley Additive exPlanations~\cite{Lundberg2017} (\texttt{TreeExplainer}); global importance is reported as the mean absolute SHAP value $\overline{|\phi_j|}=N^{-1}\sum_i|\phi_{ij}|$ on the transformed model-output scale and is therefore dimensionless. The workflow is summarized in Fig.~\ref{fig:ml_workflow}.

\section{Results}
\label{sec:results}

\subsection{Machine-learning model performance}

\begin{figure}[t]
 \centering
\includegraphics[width=0.4\textwidth]{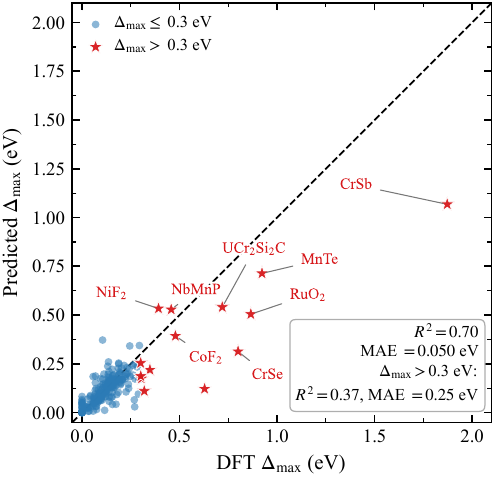}
  \caption{
    Five-fold cross-validated parity plot of the machine-learned
    $\Delta_\mathrm{max}$ against the DFT reference for the $212$ training
    materials. Circles: small splitters ($\Delta_\mathrm{max}\leq0.3$~eV);
    stars: large splitters ($\Delta_\mathrm{max}>0.3$~eV). The dashed line
    is the ideal one-to-one relation, and selected altermagnets are
    labelled. Inset: cross-validated $R^2$ and mean absolute error, overall
    and for the large-splitter subset.
  }
  \label{fig:ml_parity}
\end{figure}

We first assess the model's predictive performance and then examine the feature attributions and descriptor, property trends used to formulate candidate-prioritization criteria.
Figure~\ref{fig:ml_parity} shows the 5-fold cross-validation parity plot of DFT-computed versus ML-predicted $\Delta_\mathrm{max}$. The 180 materials naturally divide into two populations: 166 small-splitter materials ($\Delta_\mathrm{max}\leq 0.3$~eV, blue circles) and 14 large-splitter materials ($\Delta_\mathrm{max}>0.3$~eV, red stars), reflecting the heavy class imbalance of the training set in which only 7.7\% of materials exhibit large splittings. The model achieves an overall cross-validation $R^2=0.7$ and mean absolute error $\mathrm{MAE}=0.050$~eV. Performance on the scientifically important large-splitter subset is lower ($R^2=0.368$, $\mathrm{MAE}=0.248$~eV), as expected from training-set imbalance, yet the model correctly ranks all 14 large-splitter materials above the median prediction and reproduces the ordering among them. Several well-known altermagnets are annotated directly on the plot; the model reproduces CrSb ($\Delta_\mathrm{max}^{\mathrm{DFT}}=1.87$~eV) as the material with the highest predicted splitting in the training set, confirming that the surrogate has learned the physical factors that drive extreme splittings even when they are statistically rare.

\subsection{Feature importance and design principles}
\label{sec:shap}

\begin{figure}[htbp]
 \centering
\includegraphics[width=0.5\textwidth]{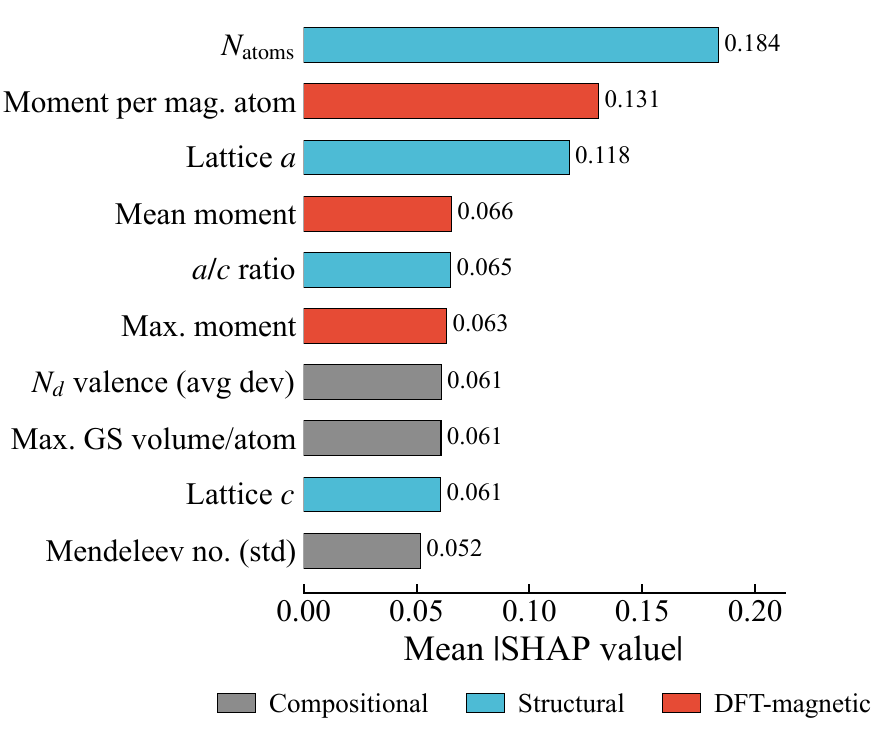}
 \caption{%
    Global feature importance of the gradient-boosted surrogate, quantified
    as the mean absolute SHAP value
    $\overline{|\phi_j|}=N^{-1}\sum_i|\phi_{ij}|$ over the
    $N=180$ training materials. The ten highest-ranked of the 50 retained
    descriptors are shown, coloured by descriptor class: compositional
    (Magpie elemental statistics), structural (unit-cell geometry and
    symmetry), and DFT-magnetic (site-resolved moments and the spin-ordering
    angle).
  }
  \label{fig:ml_shap}
\end{figure}

Figure~\ref{fig:ml_shap} shows the ten descriptors with the largest mean absolute SHAP values on the log-transformed model-output scale. The number of atoms per unit cell, $N_\mathrm{atoms}$, has the largest model attribution, $\overline{|\phi|}=0.184$, followed by the magnetic moment per magnetic atom, $\overline{|\phi|}=0.131$, and the in-plane lattice parameter $a$, $\overline{|\phi|}=0.118$. Summed over all 50 retained descriptors, compositional, structural, and DFT-derived magnetic descriptors account for 38.9\%, 33.5\%, and 27.6\%, respectively, of the total mean absolute attribution. The model therefore relies predominantly on structural and compositional information, while the DFT-derived magnetic descriptors provide a separate and quantitatively significant contribution. These attributions describe how the surrogate distributes its predictive weight and should not, by themselves, be interpreted as causal relationships.

\begin{figure}[t]
 \centering
\includegraphics[width=0.4\textwidth]{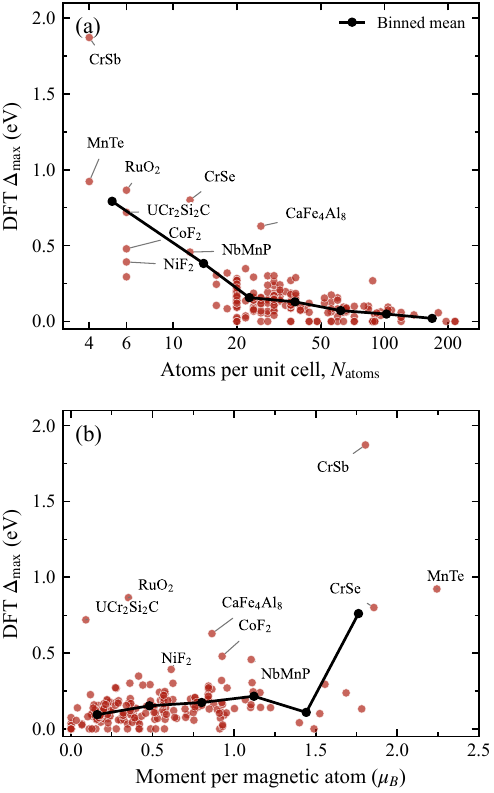}
 \caption{%
    Dependence of the DFT spin splitting $\Delta_\mathrm{max}$ on the two
    descriptors ranked highest by the SHAP analysis for the training
    materials. (a)~$\Delta_\mathrm{max}$ against the number of atoms per
    unit cell, on a logarithmic scale; (b)~$\Delta_\mathrm{max}$ against the
    magnetic moment per magnetic atom. Black lines connect the mean of
    $\Delta_\mathrm{max}$ within each bin, bins containing fewer than three
    materials being omitted. Selected altermagnets are labelled. 
  }
  \label{fig:ml_trends}
\end{figure}

To examine the direction of the two strongest associations, Fig.~\ref{fig:ml_trends} plots the DFT-computed splitting against $N_\mathrm{atoms}$ and the magnetic moment per magnetic atom.
Figure~\ref{fig:ml_trends}(a) shows that the largest splittings are concentrated among materials with small unit cells and that the binned mean generally decreases with increasing $N_\mathrm{atoms}$.
Figure~\ref{fig:ml_trends}(b) reveals a positive overall association between splitting and local moment, as expected theoretically~\cite{Roig24}, although the relation is scattered and nonmonotonic. A large moment therefore favors, but neither
guarantees nor is required for, a large splitting.

Together, the SHAP attributions and the underlying data suggest two candidate-prioritization criteria: compact primitive cells and magnetic sublattices capable of sustaining sizable local moments. These criteria are applied only after the symmetry conditions for altermagnetism have been satisfied; they rank symmetry-allowed candidates rather than determine whether altermagnetism is present.

\subsection{High-throughput screening of \texorpdfstring{$A_2XY$}{A2XY} Heusler compounds}
\label{sec:a2xy}

\begin{figure*}[t]
  \centering
  \includegraphics[width=\textwidth, trim=0pt 25pt 0pt 0pt, clip]{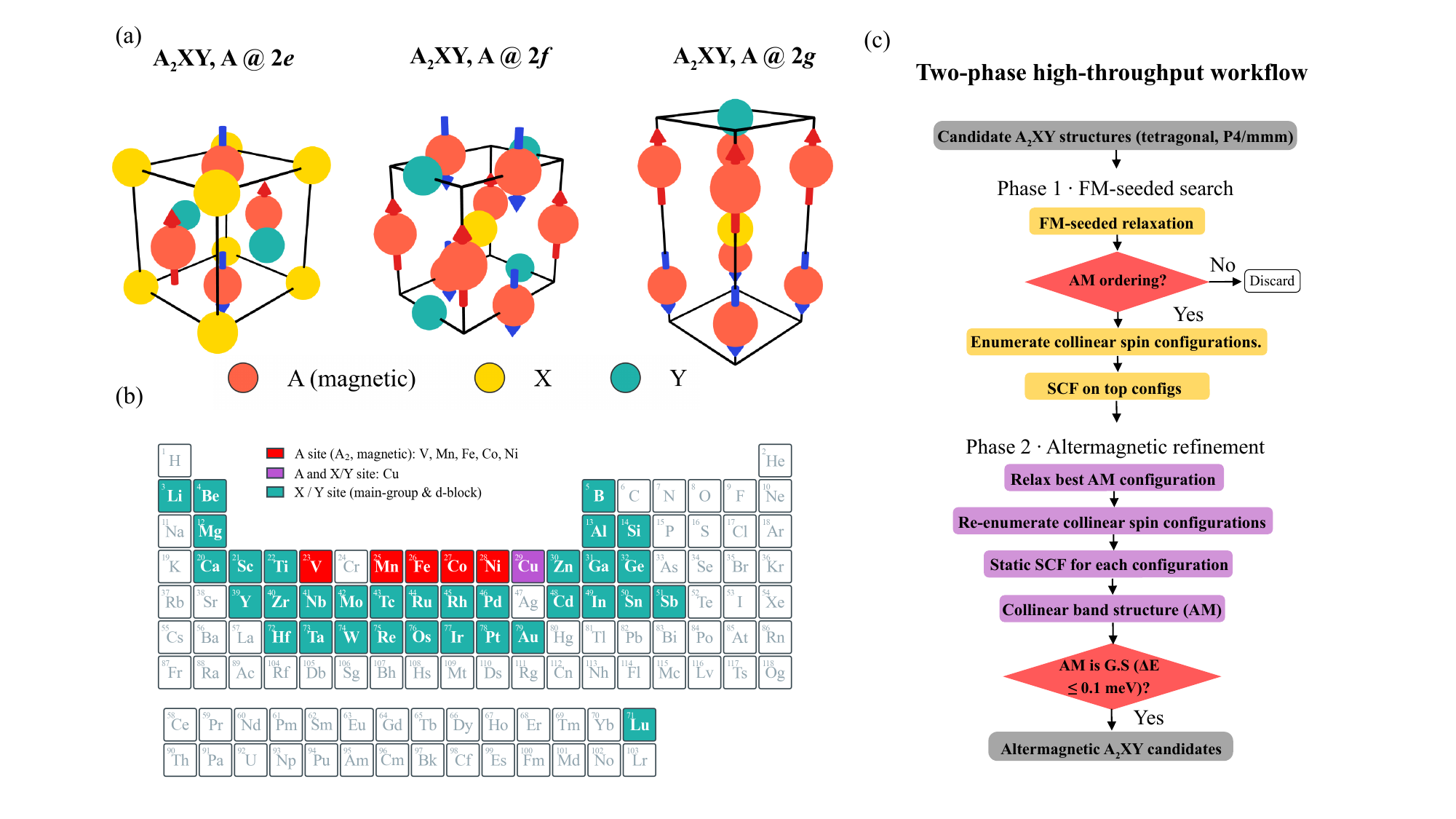}
  \caption{%
(a)~Three representative $P4/mmm$ (No.~123) arrangements, with the $A$ atoms occupying the $2e$ (left), $2f$ (center), or $2g$ (right) Wyckoff orbit and the $X$ and $Y$ atoms occupying one-fold sites. The $2e$ and $2f$ arrangements are related by an origin shift along $c$. Red and blue arrows denote the compensated antiparallel moments along $c$.
(b)~Elements sampled by crystallographic role: magnetic $A$-site elements (V, Mn, Fe, Co, and Ni), $X/Y$-site elements (main-group and $d$-block), and Cu, which was considered in both roles.
(c)~Two-phase DFT workflow. Phase~1 relaxes a ferromagnetic seed, removes structures that fail to stabilize a magnetic moment or the required antiparallel configuration, and ranks the symmetry-distinct collinear spin configurations by SCF energy. Phase~2 fully relaxes the lowest-energy AM configuration, re-enumerates the collinear orderings, and calculates the AM band structure and maximum splitting $\Delta*\mathrm{max}$. A candidate is retained when the AM ordering is the lowest-energy tested collinear configuration or lies within $\Delta E_{\mathrm{AM\text{-}GS}}\leq0.1$~meV/f.u.\ of it.
}

  \label{fig:a2xy_comp}
\end{figure*}

To test both criteria mentioned above in a distinct chemical family, we focus on ternary $A_2XY$ Heusler intermetallics [Fig.~\ref{fig:a2xy_comp}(a)]. Their four-atom primitive cells address the first criterion, while the two transition-metal $A$ sites provide a natural route to sizable local moments, which are subsequently verified by DFT. We retrieved 335 $A_2XY$ structures from the Alexandria database~\cite{alexandria} and calculated their magnetic order and altermagnetic band splitting directly by DFT. Of these structures, 307 retain the simple-tetragonal space group $P4/mmm$ (No.~123) after relaxation and constitute the screening set analyzed here; the remaining 28 relax to lower-symmetry structures and were excluded.

Within this family, whether a given compound realizes altermagnetism at all depends on the crystal symmetry, specifically by where the magnetic atoms sit in the lattice. The $A_{2}XY$ compounds contain a pair of magnetic transition-metal atoms $A$ (\textit{A}\,$=$\,V, Cr, Mn, Fe, Co, Ni), Cu in both A and X/Y roles, and two non-magnetic ligand atoms, $X$ and $Y$. As summarized in Fig.~\ref{fig:a2xy_comp}(b), the ligand sites are chemically diverse: each may host either a main-group ($s/p$-block) element or a non-magnetic ($d$-block) transition metal, and in nearly half of the compositions both ligands are transition metals. The paramagnetic point group is $4/mmm$ ($D_{4h}$, order 16), generated by the fourfold rotation $C_{4z}$, a vertical mirror $m_{x}$, and spatial inversion $\mathcal{P}$; the remaining symmetry operations follow by composition of these three. The non-magnetic $X$ and $Y$ atoms each occupy one of the maximal-symmetry Wyckoff positions $1a$, $1b$, $1c$, or $1d$ (site symmetry $4/mmm$), leaving the two magnetic $A$ atoms to share a single multiplicity-two orbit. Within this structural family, the Wyckoff orbit of the $A$ pair determines whether the imposed compensated ordering has altermagnetic or conventional-antiferromagnetic symmetry, whereas chemistry determines its energetic and structural stability.

Across the family the magnetic $A_{2}$ pair occupies one of the four multiplicity-two Wyckoff orbits of $P4/mmm$, which fall into two symmetry classes; three representative configurations are illustrated in Fig.~\ref{fig:a2xy_comp}(a). In two of them the $A$ atoms lie in a plane normal to $c$, occupying either the edge-centered $2f$ sites $(0,\tfrac{1}{2},0),(\tfrac{1}{2},0,0)$ or the $2e$ sites $(0,\tfrac{1}{2},\tfrac{1}{2}),(\tfrac{1}{2},0,\tfrac{1}{2})$ (site symmetry $mmm$); in both, the two atoms are mapped onto one another by the fourfold rotation $C_{4z}$. In the third, the $A$ atoms sit on the fourfold axis at the $2g$ sites $(0,0,\pm z)$ (site symmetry $4mm$) and are instead related by inversion $\mathcal{P}$; the remaining on-axis orbit, $2h$ at $(\tfrac{1}{2},\tfrac{1}{2},\pm z)$, occurs rarely and belongs to the same class as $2g$. All four retain the full $P4/mmm$ symmetry and differ only in the operation relating the two magnetic sublattices.

\begin{figure}[t]
  \centering
  \includegraphics[width=0.5\textwidth, trim=0pt 20pt 0pt 0pt]{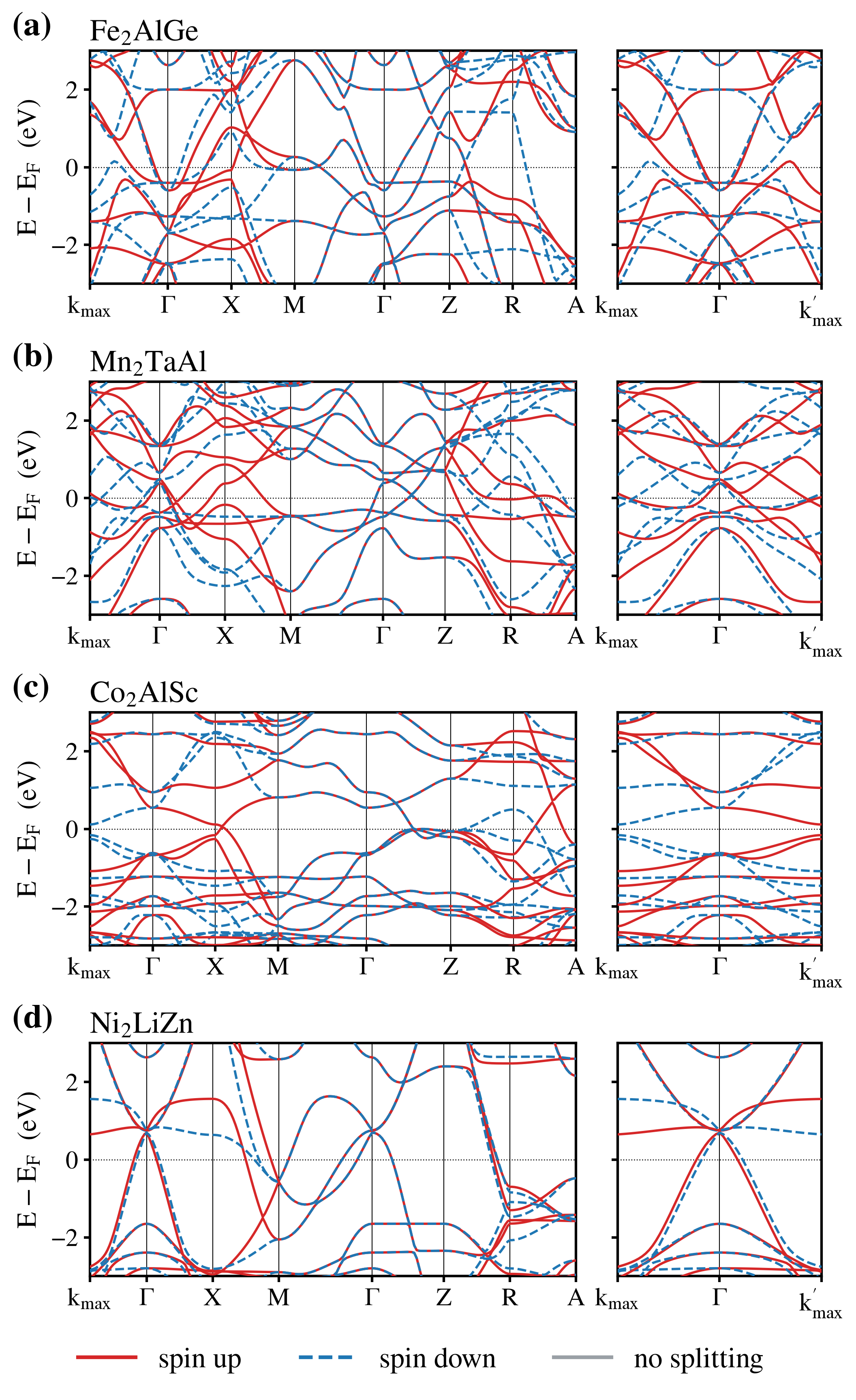}
  \caption{Spin-resolved band structures of the altermagnetic $A_2XY$ candidates
(a)~Fe$_2$AlGe, (b)~Mn$_2$AlTa, (c)~Co$_2$AlSc, and (d)~Ni$_2$LiZn, computed
without spin-orbit coupling. In each panel the left subplot follows the
high-symmetry path $\Gamma$--X--M--$\Gamma$--Z--R--A of the $P4/mmm$ Brillouin
zone, anchored at $k_\mathrm{max}$, the momentum of maximum spin splitting; the
right subplot spans $k_\mathrm{max}$--$\Gamma$--$k'_\mathrm{max}$, with
$k'_\mathrm{max}$ the $C_{4z}$ image of $k_\mathrm{max}$. Red (solid) and blue
(dashed) lines denote spin-up and spin-down bands; grey lines are spin-degenerate
($|E_\uparrow - E_\downarrow| < 1$~meV). Energies are referenced to
$E_\mathrm{F}$. The spin splitting reverses sign between $k_\mathrm{max}$ and
$k'_\mathrm{max}$, the hallmark of altermagnetism.}
\label{fig:a2xy_bands}
\end{figure}

When the sublattices order antiparallel, this connecting operation acquires a $\pi$ spin rotation, and its character fixes the spin symmetry of the electronic structure~\cite{Smejkal2022a,Smejkal2022b}. For the on-axis $2g$ ($2h$) configuration the two $A$ atoms are inversion partners, so the ordered state preserves the combined symmetry $\mathcal{P}\mathcal{T}$, where $\mathcal{T}$ is time reversal. This enforces a Kramers-like twofold spin degeneracy at every momentum, and the compound is a conventional, fully compensated collinear antiferromagnet. For the in-plane $2e$ and $2f$ configurations each $A$ atom instead lies at its own inversion center, so $\mathcal{P}$ maps each sublattice onto itself rather than onto the other. Consequently, neither $\mathcal{P}\mathcal{T}$ nor $\{\mathcal{T}\,|\,\boldsymbol{\tau}\}$ is a symmetry of the ordered state. Instead, the opposite-spin sublattices are related by the fourfold spatial rotation combined with spin reversal, written $[C_2\,||\,C_{4z}]$ in spin-space-group notation~\cite{Litvin1977,Chen2024}. This symmetry permits momentum-dependent spin splitting while maintaining zero net magnetization. The splitting changes sign under $C_{4z}$ because the rotation exchanges the two spin sublattices, whereas inversion preserves each sublattice and therefore gives $\Delta(-\mathbf{k})=\Delta(\mathbf{k})$. The resulting even-parity $d$-wave momentum dependence is the characteristic altermagnetic form for the $2e$ and $2f$ configurations~\cite{Smejkal2022a,Smejkal2022b}; the $2g$ and $2h$ configurations instead retain spin degeneracy through $\mathcal{P}\mathcal{T}$.

We classified every candidate by this criterion with the \textsc{amcheck} package~\cite{amcheck}, and independently confirmed each verdict from the crystallographic symmetry operations returned by \textsc{spglib}~\cite{spglib}; the two approaches agree for all structures. Of the 307 candidates in $P4/mmm$, 169 adopt one of the two altermagnetic arrangements ($2f$ or $2e$) and 138 the conventional antiferromagnetic one ($2g$, together with its infrequent on-axis variant $2h$), spanning all six magnetic species. A substantial fraction of compositions appear as both an altermagnetic and a conventional-antiferromagnetic polymorph, confirming that the altermagnetic phase is selected by the geometry of the $A$ sublattice rather than by chemistry alone. The four compounds examined in detail below, Fe$_{2}$AlGe, Mn$_{2}$AlTa, Co$_{2}$AlSc, and Ni$_{2}$LiZn, all crystallize in the altermagnetic $2f$ configuration.

The two-phase workflow is summarized in Fig.~\ref{fig:a2xy_comp}(c). In Phase~1, each $P4/mmm$ structure is relaxed from a ferromagnetic seed. Structures that fail to stabilize a magnetic moment or the antiparallel configuration required for altermagnetism are removed, after which the symmetry-distinct collinear configurations are enumerated and ranked by self-consistent total energy. In Phase~2, the lowest-energy altermagnetic configuration is fully relaxed, the collinear orderings are re-enumerated on the resulting structure, and each ordering is evaluated by a static self-consistent calculation. The ferromagnetic reference retains its separately relaxed Phase-1 geometry, whereas the altermagnetic and other antiferromagnetic energies are evaluated on the altermagnetically relaxed Phase-2 geometry.

\begin{figure}[t]
  \centering
  \includegraphics[width=\columnwidth]{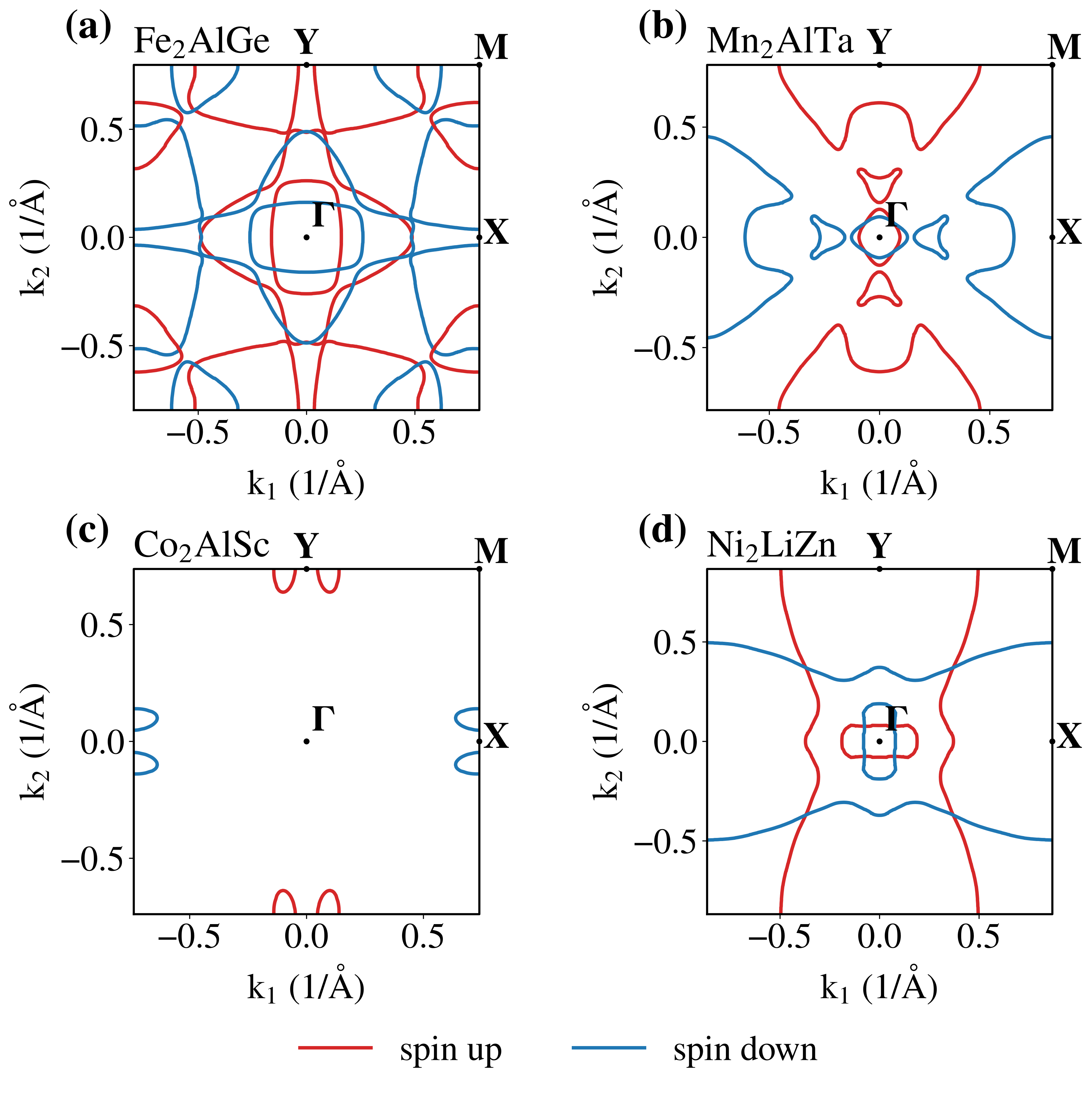}
\caption{Calculated Fermi surfaces of the four $A_2XY$ altermagnetic
candidates in the $k_1$--$k_2$ plane through $\Gamma$ perpendicular to the
crystallographic four-fold ($C_4$) axis of space group $P4/mmm$ (No.\,123).
Red and blue contours mark the spin-up and spin-down Fermi sheets,
respectively.}
\label{fig:a2xy_fs2d}
\end{figure}

Of the 169 symmetry-allowed structures, 154 stabilize as metallic altermagnets in DFT. Sixteen have an altermagnetic state that is either the lowest-energy tested collinear configuration or lies within the operational tolerance $\Delta E_{\mathrm{AM\text{-}GS}}\leq0.1$~meV/f.u.; these compounds are listed in Table~\ref{tab:a2xy}. This tolerance groups numerically indistinguishable solutions and should not be interpreted as sub-meV DFT accuracy. Spin-resolved band structures, relaxed coordinates, site-resolved magnetic moments, and Brillouin-zone paths for all 154 DFT-confirmed candidates are provided in Fig.~S3 of the Supplemental Material.

The 16 retained compounds were subsequently evaluated for dynamical and mechanical stability. Fifteen have no imaginary phonon modes and satisfy the Born elastic-stability criteria. V$_2$MoW is the sole exception: it exhibits a soft mode near $\Gamma$ ($\omega_\mathrm{min}\approx-5\ \mathrm{cm}^{-1}$) and violates the mechanical-stability criterion through $C_{66}=-11$~GPa. The agreement between the phonon and elastic tests supports its exclusion from the subsequent response analysis, leaving 15 dynamically and mechanically stable candidates (see Table S2 for details). 

Thermodynamic stability was assessed within GGA$+U$ using competing phases drawn independently from the Materials Project and Alexandria databases. Table~\ref{tab:a2xy} reports the Materials-Project-based results, which we adopt as the primary standardized reference, while the complete results obtained from both competing-phase sets are provided in Tables~S3 and S4. We define the signed stability energy as $\Delta E_\mathrm{hull}^{+U}=E_{P4/mmm}-E_\mathrm{hull}^{\mathrm{MP}}$, where $E_\mathrm{hull}^{\mathrm{MP}}$ is the energy of the competing-phase hull constructed from Materials Project entries. Negative values place the tetragonal phase below this hull, whereas positive values place it above.

Of the 15 dynamically and mechanically stable candidates, 10 have negative Materials-Project-based values. These include three of the four compounds examined in detail: Fe$_2$AlGe, Mn$_2$AlTa, and Co$_2$AlSc, with $\Delta E_\mathrm{hull}^{+U}=-72$, $-53$, and $-30$~meV/atom, respectively. The remaining five compounds, Co$_2$IrNb, Ni$_2$LiZn, V$_2$RuW, Co$_2$TiZn, and Co$_2$BeTi, have positive values ranging from $7.5$ to $54.1$~meV/atom. For Ni$_2$LiZn, Co$_2$TiZn, and Co$_2$BeTi, the lower-energy competitor is a cubic polymorph of the same composition, whereas Co$_2$IrNb and V$_2$RuW lie above multicomponent decomposition mixtures. Although all candidates and competing phases were fully relaxed and evaluated using identical computational settings, hull distances within a few tens of meV/atom should be regarded as near-degenerate because they remain sensitive to the set of competing phases included and to the intrinsic accuracy of the GGA$+U$ approximation. Consistently, the independent Alexandria-based analysis shifts several individual hull distances but preserves the overall conclusion that most candidates are stable or close to the competing-phase hull.

The $A_2XY$ compounds exhibit exceptionally large splittings along the sampled band paths (Table~\ref{tab:a2xy}). The maximum sampled splitting $\Delta_\mathrm{max}$ reaches $2.24$~eV in Co$_2$AlSc and $2.07$~eV in Fe$_2$AlGe. Six candidates exceed the $1.43$~eV obtained for CrSb using the same computational settings and sampling protocol~\cite{Ding2024,Zhang2025}. The path-averaged splitting $\bar{\Delta}$ is largest in Mn$_2$AlTa ($0.98$~eV), showing that its large splitting extends over the sampled band path rather than being confined to a single momentum.

\begin{figure*}[t]
  \centering
  \includegraphics[width=0.9\textwidth, trim=0pt 70pt 0pt 0pt]{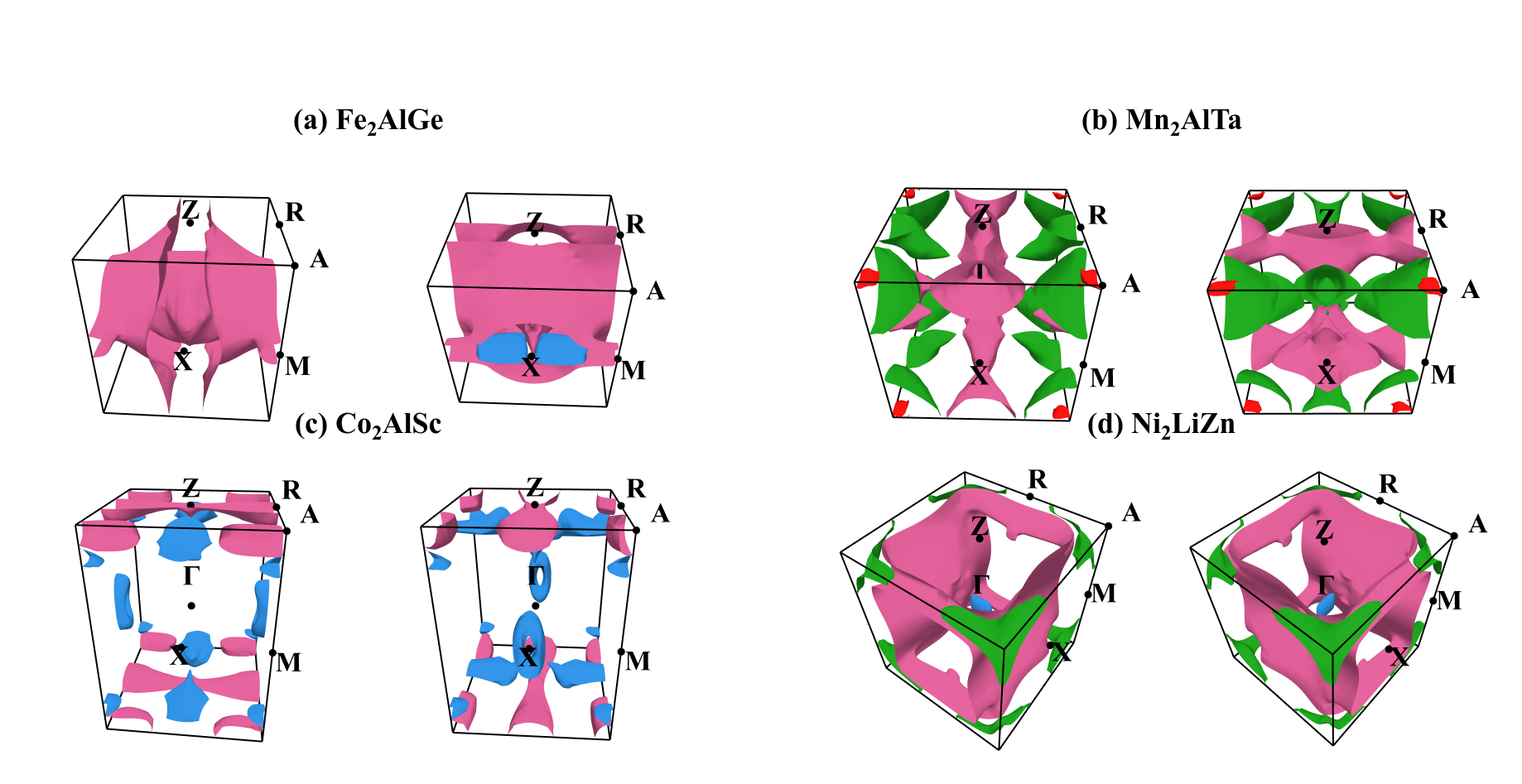}
\caption{Calculated three-dimensional Fermi surfaces:
(a)~Fe$_{2}$AlGe, (b)~Mn$_{2}$AlTa, (c)~Co$_{2}$AlSc, and (d)~Ni$_{2}$LiZn.
For each material, the left panel shows the spin-up Fermi surface and the
right panel the spin-down Fermi surface within the first Brillouin zone. The spin-up and spin-down sheets are
manifestly non-degenerate throughout the zone, and their shapes are related
by a fourfold rotation about the crystallographic $c$ axis combined with
time reversal, the momentum-space signature of $d$-wave altermagnetic spin
splitting.}
\label{fig:a2xy_fs3d}
\end{figure*}

\begin{table*}[t]

\caption{%
Altermagnetic $A_2XY$ candidates from the Alexandria database with $\Delta E_{\mathrm{AM\text{-}GS}} \leq 0.1$\,meV/f.u. $\Delta E_{\mathrm{AM\text{-}GS}}$: energy of the altermagnetic configuration above the overall collinear ground state (Phase~2). $\Delta_{\mathrm{max}}$, $\bar{\Delta}$: maximum and mean collinear spin splitting over the uniform $k$-mesh of the static Phase~2 calculation. $\Delta E_\mathrm{hull}^{+U}=E_{P4/mmm}-E_\mathrm{hull}^{\mathrm{MP}}$ is the signed energy relative to the GGA$+U$ competing-phase hull constructed from Materials Project entries; negative and positive values place the tetragonal phase below and above this hull, respectively. ``Dyn./Mech.'': dynamical (phonon) and mechanical (Born) stability (\checkmark\ both satisfied, $\times$ violated). Sorted by $\Delta_{\mathrm{max}}$; Alexandria IDs are given without the \texttt{agm} prefix.
  } 
\label{tab:a2xy}
\centering
\footnotesize
\setlength{\tabcolsep}{6pt}
\renewcommand{\arraystretch}{0.95}
\begin{tabular}{l l r r r r c}
\toprule
Formula & \multicolumn{1}{c}{Alexandria ID}
        & $\Delta E_{\mathrm{AM\text{-}GS}}$
        & $\Delta_{\mathrm{max}}$
        & $\bar{\Delta}$
        & $\Delta E_{\mathrm{hull}}^{+U}$
        & Dyn./Mech. \\
        &
        & (meV) & (eV) & (eV) & (meV/atom) & \\
\midrule
Co$_2$AlSc & 001197600 & 0.078 & 2.235 & 0.813 & $-30.3$ & \checkmark \\
Fe$_2$AlGe & 001196968 & 0.000 & 2.070 & 0.784 & $-72.0$ & \checkmark \\
Mn$_2$AlTa & 003188011 & 0.084 & 1.604 & 0.976 & $-53.0$ & \checkmark \\
Co$_2$GeZn & 001202417 & 0.000 & 1.585 & 0.544 & $-8.0$  & \checkmark \\
Co$_2$TiZn & 001228663 & 0.096 & 1.507 & 0.679 & $+51.3$ & \checkmark \\
Fe$_2$GaTi & 001228742 & 0.000 & 1.444 & 0.646 & $-65.0$ & \checkmark \\
Co$_2$BeTi & 001199021 & 0.000 & 1.395 & 0.665 & $+54.1$ & \checkmark \\
Co$_2$IrNb & 001202277 & 0.000 & 1.228 & 0.591 & $+7.5$  & \checkmark \\
Ni$_2$LiZn & 003192666 & 0.000 & 1.006 & 0.556 & $+19.2$ & \checkmark \\
V$_2$PdRh  & 001229872 & 0.000 & 0.813 & 0.449 & $-147.7$ & \checkmark \\
Co$_2$BeGa & 001198799 & 0.000 & 0.606 & 0.213 & $-36.4$ & \checkmark \\
V$_2$MoW   & 001230558 & 0.000 & 0.568 & 0.202 & $-5.4$  & $\times$ \\
V$_2$IrPt  & 001229757 & 0.000 & 0.563 & 0.344 & $-78.2$ & \checkmark \\
V$_2$RuW   & 001229890 & 0.000 & 0.514 & 0.195 & $+20.4$ & \checkmark \\
V$_2$IrRh  & 001229868 & 0.000 & 0.430 & 0.227 & $-100.3$ & \checkmark \\
Co$_2$AlGa & 001196837 & 0.000 & 0.414 & 0.206 & $-44.5$ & \checkmark \\
\bottomrule
\end{tabular}
\renewcommand{\arraystretch}{1.0}
\end{table*}

These results directly corroborate the design principles extracted by our machine-learning model. The $A_2XY$ Heuslers realize both rules at once: a compact four-atom unit cell and $3d$ transition-metal sublattices carrying large local moments. That precisely this structural and chemical combination produces the largest splittings in the screen, surpassing CrSb, confirms that the compact-cell and large-moment criteria learned from the MAGNDATA training set are genuine, transferable predictors of strong altermagnetic splitting.

The momentum-dependent spin splittings are shown in Fig.~\ref{fig:a2xy_bands}. In each compound, the antiparallel sublattice moments yield zero net magnetization, while the electronic bands are strongly spin-split over extended regions of the Brillouin zone. Along high-symmetry directions contained within the symmetry-protected nodal planes, the spin-up and spin-down bands remain degenerate (gray, $|E_\uparrow-E_\downarrow|<1$~meV), whereas away from these planes they separate by as much as $2.24$~eV. This behavior differs from ferromagnetic exchange splitting, which is accompanied by a net magnetization, and from spin-orbit-driven splitting in noncentrosymmetric systems. Here, the splitting occurs without spin-orbit coupling, is even under inversion, and coexists with zero net magnetization.

The $d$-wave altermagnetic symmetry is demonstrated explicitly in the right subplots, which follow the path $k_\mathrm{max}$--$\Gamma$--$k'_\mathrm{max}$, where $k'_\text{max}$ is obtained from $k_\text{max}$ via a $90^\circ$ rotation around the $z$-axis. The spin-up and spin-down bands interchange between these two momenta. In all four compounds, spin-split bands cross $E_\mathrm{F}$, making the splitting relevant to Fermi-surface transport and accessible to spectroscopic probes.

\begin{table*}[t]
\centering
\caption{Fermi-surface spin polarization $\langle|P|\rangle$ and Julliere
tunneling magnetoresistance (TMR
$=2\langle|P|\rangle^{2}/(1-\langle|P|\rangle^{2})$) at $E_\mathrm{F}$, for
transport along $[100]$, the in-plane diagonal, and $[001]$. For the
tetragonal compounds $[010]$ is equivalent to $[100]$ and is not tabulated.
The nodal structures of the different altermagnetic orders are directly
visible: in the $A_2XY$ compounds ($d_{x^2-y^2}$) the $\{110\}$ planes are
nodal and transport along $[110]$ is spin-unpolarized, verified numerically
($\langle|P|\rangle\leq 0.003$) for fourteen of the fifteen compounds; in
rutile MnF$_2$ ($d_{xy}$) the $\{100\}$ planes are nodal instead; in the
hexagonal $g$-wave compounds the polarized channels are basal, with $[001]$
and the $[1\bar{1}0]$ family suppressed.}
\label{tab:fs_tmr}
\setlength{\tabcolsep}{9pt}
\renewcommand{\arraystretch}{1.05}
\begin{tabular}{l cc cc cc}
\toprule
& \multicolumn{2}{c}{$[100]$} & \multicolumn{2}{c}{$[110]$\footnotemark[1]}
& \multicolumn{2}{c}{$[001]$} \\
\cmidrule(lr){2-3}\cmidrule(lr){4-5}\cmidrule(lr){6-7}
Material & $\langle|P|\rangle$ & TMR (\%) & $\langle|P|\rangle$ & TMR (\%)
         & $\langle|P|\rangle$ & TMR (\%) \\
\midrule
Co$_2$AlGa & 0.20 & 8   & 0.00 & 0 & 0.15 & 5   \\
Co$_2$AlSc & 0.69 & 182 & 0.00 & 0 & 0.60 & 113 \\
Co$_2$BeGa & 0.38 & 34  & 0.00 & 0 & 0.24 & 12  \\
Co$_2$BeTi & 0.66 & 154 & 0.00 & 0 & 0.63 & 132 \\
Co$_2$GeZn & 0.56 & 91  & 0.00 & 0 & 0.53 & 78  \\
Co$_2$IrNb & 0.53 & 78  & 0.00 & 0 & 0.27 & 16  \\
Co$_2$TiZn & 0.80 & 356 & 0.00 & 0 & 0.49 & 63  \\
Fe$_2$AlGe & 0.50 & 67  & 0.00 & 0 & 0.44 & 48  \\
Fe$_2$GaTi\footnotemark[2] & 0.61 & 119 & 0.12 & 3 & 0.74 & 242 \\
Mn$_2$AlTa & 0.47 & 57  & 0.00 & 0 & 0.50 & 67  \\
Ni$_2$LiZn & 0.66 & 154 & 0.00 & 0 & 0.56 & 91  \\
V$_2$IrPt  & 0.43 & 45  & 0.00 & 0 & 0.25 & 13  \\
V$_2$IrRh  & 0.26 & 15  & 0.00 & 0 & 0.21 & 9   \\
V$_2$PdRh  & 0.33 & 24  & 0.00 & 0 & 0.28 & 17  \\
V$_2$RuW   & 0.41 & 40  & 0.00 & 0 & 0.51 & 70  \\
\midrule
\multicolumn{7}{l}{\textit{Reference $d_{xy}$ altermagnet (rutile, same method)}}\\
MnF$_2$\footnotemark[3] & 0.01 & 0 & 0.83 & 443 & 0.38 & 34 \\
\midrule
\multicolumn{7}{l}{\textit{Reference $g$-wave altermagnets (hexagonal, same method)\footnotemark[1]}}\\
CrSb           & 0.52 & 74  & 0.06 & 1  & 0.00 & 0 \\
MnTe\footnotemark[3] & 0.64 & 139 & 0.30 & 20 & 0.02 & 0 \\
CoNb$_4$Se$_8$ & 0.43 & 45  & 0.01 & 0  & 0.00 & 0 \\
\bottomrule
\end{tabular}
\footnotetext[1]{For the hexagonal compounds the $[110]$ column lists the
$[1\bar{1}0]$ ($\mathbf{a}-\mathbf{b}$) direction, which is not
symmetry-equivalent to $[100]$; the $[110]$ ($\mathbf{a}+\mathbf{b}$)
direction is equivalent to $[100]$ by the sixfold rotation and is not
tabulated separately.}
\footnotetext[2]{Fe$_2$GaTi shows a finite residual
polarization along $[110]$; its altermagnetic ordering differs from the
simple checkerboard of the remaining compounds.}
\footnotetext[3]{MnF$_2$ and MnTe are gapped at $E_\mathrm{F}$ in our
calculations; their values are evaluated at the valence-band edge nearest
$E_\mathrm{F}$ ($E-E_\mathrm{F}=-0.45$ and $-0.30$~eV, respectively) and are
band-edge rather than Fermi-surface properties.}
\end{table*}

\begin{figure}[t]
 \centering
\includegraphics[width=\columnwidth, trim=0pt 10pt 0pt 0pt]{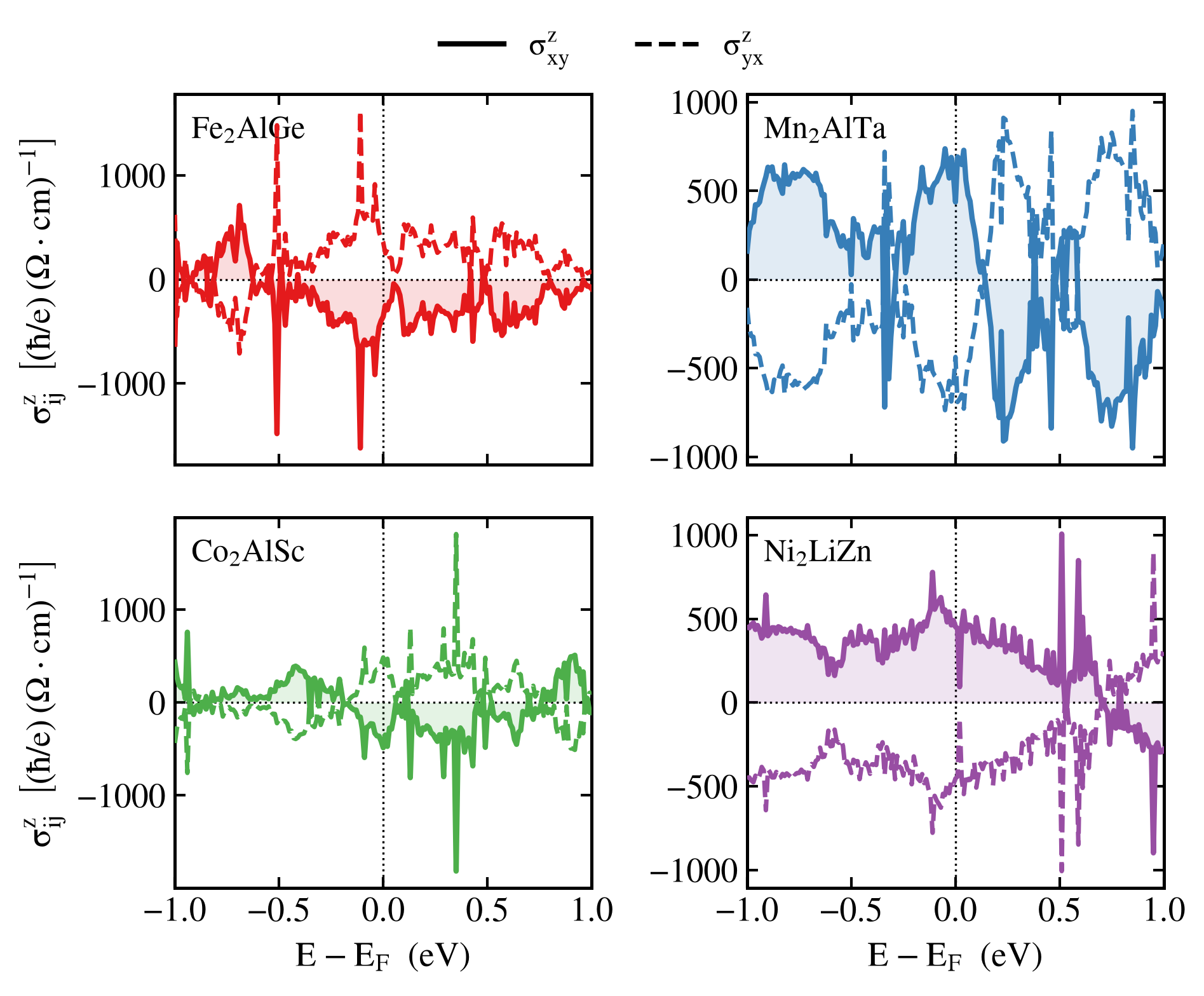}
\caption{Time-reversal-even intrinsic spin Hall conductivity $\sigma^{z}_{ij}$ as a function of the Fermi-level
position, computed within Kubo linear response from Wannier Hamiltonians
symmetrized onto the magnetic space group $P4'/mm'm$ (BNS 123.342). Solid
curves: $\sigma^{z}_{xy}$; dashed curves: $\sigma^{z}_{yx}$; shaded regions
highlight $\sigma^{z}_{xy}$. The solid and dashed curves are exact mirror images at every energy: the
diagonal mirror planes of the altermagnetic state enforce
$\sigma^{z}_{xy}=-\sigma^{z}_{yx}$. }
\label{fig:a2xy_shc}
\end{figure}

The altermagnetic momentum-space structure at the Fermi energy is further illustrated in Figs.~\ref{fig:a2xy_fs2d} and \ref{fig:a2xy_fs3d}, which show the Fermi surface cuts in the basal $k_z=0$ plane and in the full Brillouin Zone, respectively. 
In both cases, a $90^\circ$ rotation around the $z$-axis maps one spin contour onto the other. The separation between the spin-up and spin-down Fermi surfaces is generally largest along the $k_x$ and $k_y$ axes, corresponding to $[100]$ and $[010]$, and vanishes along nodal planes where $|k_x|=|k_y|$. The resulting four-lobed pattern in the $k_z=0$ plane, with spin channels interchanged by $C_{4z}$ and degeneracies retained on the nodal lines, is the characteristic Fermi-surface signature of $d_{x^2-y^2}$ altermagnetic order~\cite{Smejkal2022a,Monkman2026}.

For a transport direction $\hat{\mathbf t}$, we project the spin-resolved Fermi surfaces, calculated without spin--orbit coupling at chemical potential $\mu$ from BoltzTraP2-interpolated bands, onto the plane perpendicular to $\hat{\mathbf t}$. This plane is discretized into a $60\times60$ grid. Let $N_{\sigma}(b)$ denote the number of projected vertices from all spin-$\sigma$ Fermi-surface sheets falling within bin $b$. We define the local projected spin polarization as
  \begin{equation}
  P_{\hat{\mathbf t}}(b)=
  \frac{N_{\uparrow}(b)-N_{\downarrow}(b)}
       {N_{\uparrow}(b)+N_{\downarrow}(b)}.
  \end{equation}
Because the signed polarization cancels between symmetry-related regions of the Fermi surface, we characterize the directional spin selectivity by its mean absolute value over the set $\mathcal B$ of nonempty bins,
  \begin{equation}
  \langle|P|\rangle_{\hat{\mathbf t}}=
  \frac{1}{|\mathcal B|}
  \sum_{b\in\mathcal B}|P_{\hat{\mathbf t}}(b)|.
  \end{equation}
We use this dimensionless quantity as an effective polarization proxy in the symmetric-junction Julli\`ere estimate~\cite{Julliere1975},
  \begin{equation}
  \mathrm{TMR}=
  \frac{2\langle|P|\rangle_{\hat{\mathbf t}}^2}
       {1-\langle|P|\rangle_{\hat{\mathbf t}}^2}.
  \end{equation}

The resulting directional polarizations and TMR estimates are summarized in Table~\ref{tab:fs_tmr}. At $E_\mathrm{F}$, Co$_2$TiZn reaches $\langle|P|\rangle_{[100]}=0.80$, corresponding to an estimated TMR of $356\%$, and Co$_2$AlSc reaches $\langle|P|\rangle_{[100]}=0.69$, corresponding to $182\%$. Co$_2$BeTi and Ni$_2$LiZn each reach $154\%$ along $[100]$, while Fe$_2$GaTi reaches $242\%$ along $[001]$. These values are model estimates based on the projected Fermi surfaces rather than explicit barrier-dependent junction calculations.

For comparison, Table~\ref{tab:fs_tmr} also reports values obtained using the same procedure for MnF$_2$, CrSb, MnTe, and CoNb$_4$Se$_8$. In rutile MnF$_2$, the $d_{xy}$ order places the nodal planes on $[100]$, rotating the polarized in-plane channels by $45^\circ$ relative to those of the $A_2XY$ compounds. MnF$_2$ is insulating, however, and its listed polarization, $\langle|P|\rangle=0.83$ at $E-E_\mathrm{F}=-0.45$~eV, is a band-edge rather than a Fermi-surface value. In the metallic $g$-wave references CrSb and CoNb$_4$Se$_8$, the calculated polarization is concentrated primarily in the basal plane and is suppressed along $[001]$; semiconducting MnTe shows analogous behavior at its valence-band edge. Within this comparison set, the $A_2XY$ compounds are the only metallic systems with finite Fermi-level polarization along three mutually orthogonal principal axes, and their strongest members exceed the metallic references in the corresponding Julli\`ere estimates. Explicit barrier- and interface-dependent calculations are required to determine whether these advantages persist in experimentally realizable tunnel junctions.

Finally, we examine the Hall responses of the four representative compounds. For the N\'eel vector along the tetragonal $c$ axis, their relaxed magnetic structures belong to the magnetic space group $P4^{\prime}/mm^{\prime}m$ (BNS 123.342). Its antiunitary operation $C_{4z}\mathcal T$, the relativistic counterpart of the spin-group operation $[C_{2}\,\|\,C_{4z}]$, reverses the time-reversal-odd charge Berry curvature while mapping the Brillouin zone onto itself. It therefore forces the anomalous Hall conductivity $\sigma_{xy}$ to vanish. Because this constraint holds at every chemical potential, the anomalous Nernst response also vanishes through the Mott relation. The magnetic-group-symmetrized Wannier calculations reproduce these constraints within numerical precision. This conclusion applies to the $c$-axis N\'eel state: rotating the N\'eel vector or introducing another perturbation that breaks $C_{4z}\mathcal T$ can make an anomalous Hall response symmetry allowed~\cite{Smejkal2020}.

For spin transport, we distinguish the intrinsic time-reversal-even spin Hall conductivity calculated here, denoted $\sigma^{z}_{ij}$, from the semiclassical time-reversal-odd spin-splitter conductivity, denoted $\tilde{\sigma}^{z}_{ij}$. For spin polarization along $z$, the combined constraints of $C_{4z}\mathcal T$ and the unprimed diagonal $[110]$ mirrors require
$\sigma^{z}_{xy}=-\sigma^{z}_{yx}$ and forbid $\sigma^{z}_{xx}$, $\sigma^{z}_{yy}$, and $\sigma^{z}_{zz}$.
The spectra in Fig.~\ref{fig:a2xy_shc}, calculated from Wannier Hamiltonians symmetrized with respect to the magnetic group, satisfy these relations at every sampled energy. At $E_{F}$, $\sigma^{z}_{xy}$ equals $-356$, $+437$, $-486$, and $+453(\hbar/e)(\Omega \mathrm{cm})^{-1}$, for Fe$_2$AlGe, Mn$_2$AlTa, Co$_2$AlSc, and Ni$_2$LiZn, respectively. These magnitudes are approximately one quarter of the calculated intrinsic spin Hall conductivity of Pt reported in Ref.~\cite{Guo2008}. Within a rigid-band interpretation, $|\sigma^{z}_{xy}|$ exceeds $750 (\hbar/e)(\Omega \mathrm{cm})^{-1}$ at some chemical-potential shift within $\pm0.3$~eV for every compound, indicating substantial electronic tunability; the carrier concentrations required to reach these energies remain to be determined.

The same magnetic symmetry permits a distinct time-reversal-odd spin-splitter component satisfying $\tilde{\sigma}^{z}_{xx}=-\tilde{\sigma}^{z}_{yy}$. Consequently, for an electric field along $[110]$, this tensor form permits a spin current polarized along $z$ and flowing along the transverse $[1\bar{1}0]$ direction, even in the nonrelativistic limit~\cite{GonzalezHernandez2021,Mook2020,Jeong2026}. This response is symmetry allowed but has not been calculated here. Determining its magnitude requires a Fermi-surface transport calculation, including an explicit treatment of the relaxation time, and is left for future work.

\section{Conclusions}

We have combined interpretable machine learning with targeted high-throughput density-functional theory to identify metallic $d$-wave altermagnets in the tetragonal $A_2XY$ Heusler family. The model identifies compact primitive cells and magnetic sublattices capable of sustaining sizable local moments as candidate-prioritization criteria, while the Wyckoff orbit of the magnetic $A_2$ pair cleanly distinguishes altermagnetic from spin-degenerate conventional-antiferromagnetic arrangements within this family. DFT identifies 16 compounds in which the altermagnetic state is lowest in energy or lies within the $0.1$~meV/f.u.\ degeneracy tolerance. Fifteen satisfy the phonon and mechanical-stability criteria, and most lie below or close to the competing-phase hulls constructed independently from Materials Project and Alexandria entries. Six compounds exceed the CrSb splitting obtained using the same computational protocol, led by Co$_2$AlSc ($2.24$~eV) and Fe$_2$AlGe ($2.07$~eV). These results support the transferability of the ML-derived prioritization criteria to a chemical family outside the training set. The same symmetry produces strongly directional, multi-axis Fermi-surface responses. Within the Julliere model, Co$_2$TiZn reaches an estimated TMR of 356\% at $E_\mathrm{F}$. For a N\'eel vector along $c$, $C_{4z}\mathcal T$ forbids the anomalous Hall response, whereas the intrinsic spin Hall tensor satisfies $\sigma^z_{xy}=-\sigma^z_{yx}$ and reaches magnitudes approaching $500\,(\hbar/e)(\Omega\,\mathrm{cm})^{-1}$. The $A_2XY$ family thus provides a chemically tunable platform combining compensated magnetic order, metallic $d$-wave splitting, and symmetry-allowed spin-current responses, while demonstrating how interpretable machine learning can guide targeted first-principles discovery.  Natural next steps include the epitaxial growth of the hull-stable candidates Fe$_2$AlGe, Mn$_2$AlTa, and Co$_2$AlSc, spin-resolved photoemission of their $d$-wave splitting, and tunnel-junction and spin-torque measurements of the predicted multi-axis magnetoresistance and spin-splitter currents.

\section{Acknowledgements}
This work was partially supported by the Wallenberg Initiative Materials Science for Sustainability (WISE)
funded by the Knut and Alice Wallenberg Foundation.
R.A. acknowledges financial support from the Swedish e-Science Research Centre (SeRC).
EvL acknowledges support from the Swedish Research Council (Vetenskapsrådet, VR) under grant 2022-03090, from the Royal Physiographic Society in Lund, by NanoLund and by eSSENCE, a strategic research area for e-Science, grant number eSSENCE@LU 9:1. J.A.L. thank the Kempe-stiftelserna, Sweden and Swedish Research Council under grant no. 2023-03894 for financial support. B.M. acknowledges the financial support of Olle Engkvists stiftelse, project 207-0582. The computations were enabled by resources provided by the National Academic Infrastructure for Supercomputing in Sweden (NAISS), partially funded by the Swedish Research Council through grant agreement no. 2022-06725.

\section*{Data Availability Statement}

The data supporting the findings of this study, including relaxed structures, calculated properties, and analysis outputs, will be made available in a public repository upon publication. Additional data are available from the corresponding author upon reasonable request.

\bibliography{refs}

\end{document}